\documentclass[fleqn]{annalen}
\usepackage{epsf}
\def\vpint{\mathop{\int\hskip-0.8pc\mbox{\it /}}}
\begin{document}
\newcommand{\volume}{11}              
\newcommand{\xyear}{2000}            
\newcommand{\issue}{5}               
\newcommand{\recdate}{15 November 1999}  
\newcommand{\revdate}{dd.mm.yyyy}    
\newcommand{\revnum}{0}
\newcommand{\accdate}{dd.mm.yyyy}    
\newcommand{\coeditor}{ue}           
\newcommand{\firstpage}{000}         
\newcommand{\lastpage}{}          
\setcounter{page}{\firstpage}        
\newcommand{\keywords}{Einstein equations, integrability, integral
equations}
\newcommand{\PACS}{04.20.-q, 04.20.Jb, 04.40.Nr}
\newcommand{\shorttitle}
{G.A.Alekseev, Monodromy transform approach to solution of the Ernst
equations} 
\title{Monodromy transform approach to solution of the Ernst equations
in General Relativity}
\author{G.\ A.\ Alekseev$^{1}$}
\newcommand{\address}
  {$^{1}$Steklov Mathematical Institute of the Russian
Academy of Sciences, Gubkina str. 8,\\ \hspace*{0.5mm}  117966, GSP -
1, Moscow, Russia}

\newcommand{\email}{\tt G.A.Alekseev@mi.ras.ru}
\maketitle

\begin{abstract}
The approach, referred to as "monodromy transform", provides some
general base for solution of all known integrable space - time
symmetry reductions of Einstein equations for the case of pure vacuum
gravitational fields, in the presence of gravitationally interacting
massless fields, as well as for some string theory induced gravity
models. In this communication we present the key points of this
approach, applied to Einstein equations for vacuum and to  Einstein -
Maxwell equations for electrovacuum fields in the cases, reducible to
the known Ernst equations. Definition of the monodromy data,
formulation and solution of the direct and inverse problems of the
monodromy transform, a proof of existence and uniqueness of their
solutions, the structure of the basic linear singular integral
equations and their regularizations, which lead to the equations of
(quasi-)Fredholm type are also discussed. A construction of general
local solution of these equations is given in terms of homogeneously
convergent functional series.
\end{abstract}

\section{Introduction}
The "monodromy transform" approach \cite{GA:1985, GA:1988, GA:1993}
is called so by some analogy with the names "scattering transform" or
"spectral transform", used sometimes for the inverse scattering
methods. This approach, based on the experience of various previously
developed approaches~\footnote{Avoiding a detail citation, we refer
the readers to the references in a few papers cited here, but mainly
-- to a large and very useful F.J.Ernst's collection of related
references and abstracts, accessible throw {\sf
http://pages.slic.com/gravity}}, on the application of some powerful
ideas of the modern theory of integrable systems and on a detail
analysis of the proper internal structure of  space-time symmetry
reduced Einstein equations, provides the most general base for
solution of various {\it integrable} reductions of Einstein
equations, e.g., of vacuum and electrovacuum cases, reducible to the
known Ernst equations (in both, hyperbolic and elliptic types).
Without any change of basic elements, this approach is applicable to
solution of space-time symmetry reduced Einstein equations for
various pure gravitationally interacting massless fields
(electromagnetic and Weyl spinor fields, scalar field and stiff
matter fluid) as well as for some string theory induced gravity
models~\cite{GA:1999}.

The first step of this construction is a definition for any local
solution of the field equations of some set of functional parameters
-- {\it the monodromy data} for the corresponding solution of
associated spectral problem, which role is similar to that, playing
by the scattering data in the inverse scattering methods. (It is
remarkable, that many properties of solutions can be expressed
immediately in terms of the analytical structure of the corresponding
monodromy data and that the Einstein equations imply trivial
evolution for these data: they are functions of the spectral
parameter only, independing of space - time coordinates.) The second
step is a consideration of such transformation of "coordinates" in
the entire space of local solutions from the field potentials to the
corresponding monodromy data functions, i.e. formulation of the
{\it direct} and {\it inverse problems} of this {\it monodromy
transform}.  In particular, it turns out, that there exists a linear
singular integral equation with a scalar kernel, which solves the
mentioned above inverse problem of a reconstruction of solutions of
the Einstein equations from the monodromy data.  This equation admits
various equivalent regularizations which lead to linear integral
equations of (quasi-) Fredholm type, for which the existence and
uniqueness of solutions is proven easily for arbitrary choice of
the monodromy data. This approach and the mentioned above
integral equations, being equivalent integral equation forms of
reduced Einstein equations (in particular, of vacuum or
electrovacuum Ernst equations), admit various applications, such as
the construction of various classes of exact solutions,
which extend considerably the known classes of soliton solutions,
exact linearization of various boundary value problems -- the Cauchy
problem or characteristic initial value problem and, probably, some
others.

In this communication the key points of this construction are
described for  Einstein equations for vacuum and Einstein - Maxwell
equations for electrovacuum fields with space-time symmetries,
reducible to the Ernst equations.  Definition of the monodromy data,
formulation of the direct and inverse problems of our monodromy
transform, the structure of the basic linear singular integral
equations and their regularizations -- the equations of
(quasi-)Fredholm type are also discussed. A general local solution
of these equations is given in terms of homogeneously convergent
functional series.

\section{Direct problem of the monodromy transform}

An equivalent substitute for the study of the Ernst equations in
both, hyperbolic or elliptic cases can be the analysis of the
"spectral" problem for complex $3\times 3$
matrices
\begin{equation}\label{matrices}
{\bf \Psi}(\xi,\eta,w),\quad {\bf U}(\xi,\eta), \quad {\bf
V}(\xi,\eta), \quad  {\bf W}(\xi,\eta,w), \end{equation}
which are functions of two real or complex conjugated space-time
coordinates $\xi$, $\eta$ and a "spectral" parameter $w\in \bar C$
and which should satisfy two groups of conditions:
\begin{equation}\label{UV} \left\{\begin{array}{l}
2i(w-\xi)\partial_\xi{\bf \Psi}={\bf U}\cdot{\bf \Psi}\\[1ex]
2i(w-\eta)\partial_\eta{\bf \Psi}={\bf V}\cdot{\bf \Psi}
\end{array}\hskip1ex\right\Vert
\left.\begin{array}{l}
\mbox{rank\,}{\bf U}=1,\hskip1ex \mbox{tr\,}{\bf U}=i,\\[1ex]
\mbox{rank\,}{\bf V}=1,\hskip1ex \mbox{tr\,}{\bf V}=i.\end{array}
\hskip1ex\right\Vert\hskip1ex{\bf \Psi}(\xi_0,\eta_0,w)={\bf I}
\end{equation}
where $(\xi_0,\eta_0)$ are the coordinates of some chosen
"reference" point, and
\begin{equation}\label{integral} \left\{
\begin{array}{c}
{\bf \Psi}^\dagger \cdot{\bf W}\cdot{\bf \Psi}={\bf W}_0(w)\hfill\\
\noalign{\vspace{1ex}}
{\bf W}^\dagger_0(w)={\bf W}_0(w)\hfill
\end{array}\hskip0.5ex\right\Vert \hskip0.5ex
\displaystyle{\partial {\bf W}\over\partial w}=4i {\bf
\Omega},\hskip1ex
{\bf \Omega}=\left(\hskip-1.5ex\begin{array}{rcc} 0&1&0\\
-1&0&0\\ 0&0&0 \end{array}\hskip-0.5ex\right),\hskip1ex W^{55}=1.
\end{equation}
Here "${}^\dagger$" is a Hermitian conjugation:
${\bf \Psi}^\dagger(\xi,\eta,w)\equiv \overline{{\bf \Psi}(\xi,
\eta,\overline{w})}^T$; ${\bf W}_0(w)$ is an arbitrary Hermitian
matrix function, depending on $w$ only; the rows and columns of
$3\times 3$ - matrices are labeled by $3,4,5$, so that
$W^{55}$ is the lowest right component of ${\bf W}$.
There exists~\cite{GA:1988} a one-to-one correspondence  between
the solutions of the Ernst equations and of the spectral problem
(\ref{matrices}) - (\ref{integral}).  In particular, for the Ernst
potentials we have identifications: $\partial_\xi{\cal E}=-{\bf
U}_3{}^4$, $\partial_\eta{\cal E}=-{\bf V}_3{}^4$ and $\partial_\xi
\Phi={\bf U}_3{}^5$, $\partial_\eta \Phi={\bf V}_3{}^5$.

\begin{figure}[h] \epsfxsize=110mm
\centerline{\epsfbox{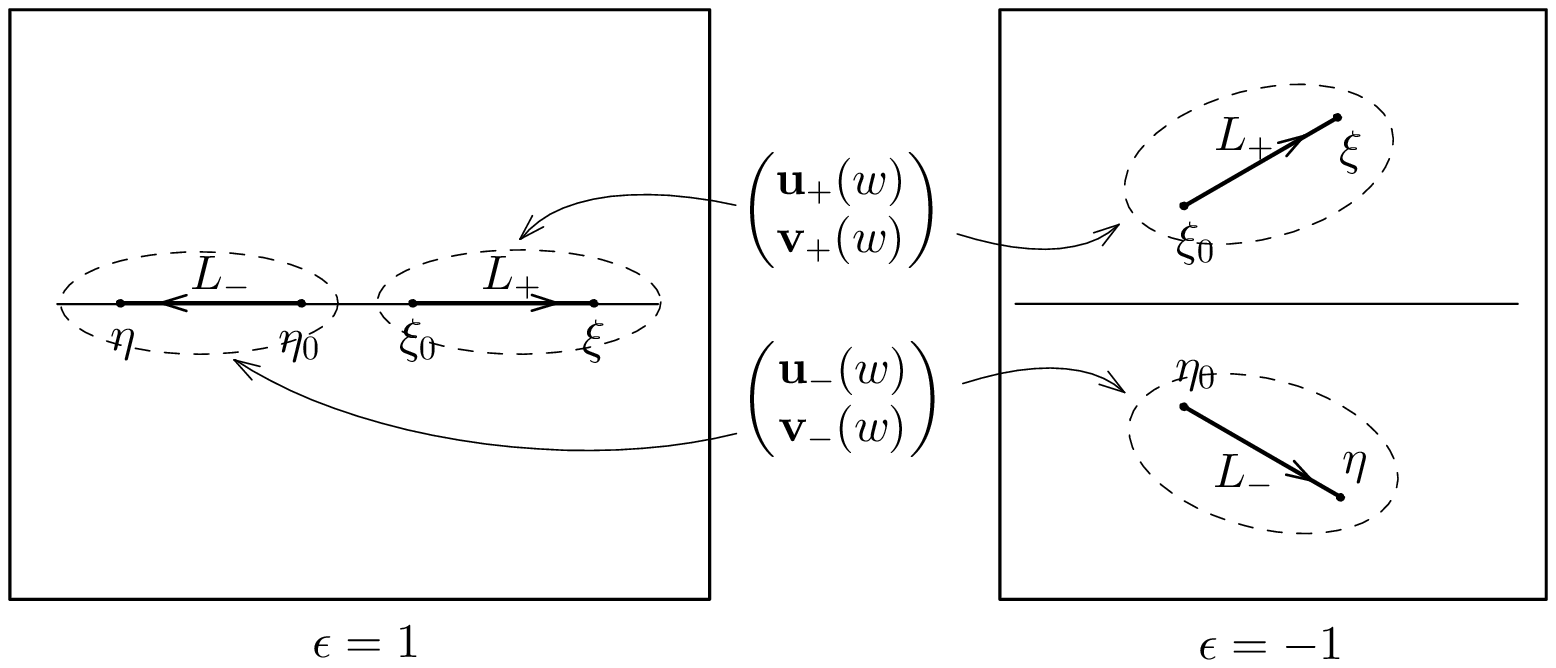}}
\caption{The singular points, the structure of the cut
$L=L_++L_-$ on the spectral plane and the domains, where the
monodromy data functions ${\bf u}_\pm(w)$ and ${\bf v}_\pm(w)$ are
defined and holomorphic, are shown here for the hyperbolic
($\epsilon=1$) and elliptic ($\epsilon=-1$) cases separately.}
\label{Fig_1} \end{figure}

A detail analysis \cite{GA:1985, GA:1988} of the analytical structure
of solutions of (\ref{matrices}) - (\ref{integral}) on the spectral
plane shows the existence of some universal properties of
${\bf \Psi}(\xi,\eta,w)$. In particular it is holomorphic function
of $w$ everywhere outside four branchpoints and the cut
$L=L_++L_-$ joining these points, as it is shown on Fig.\ \ref{Fig_1}.
Nextly it turns out, that the behaviour of ${\bf\Psi}$ near the
branchpoints can be described by so called {\it monodromy matrices}
${\bf T}_\pm (w)$, which characterise the linear transformations of
the solution ${\bf \Psi}$ of the linear system (\ref{UV}), continued
analytically along the paths $T_\pm$, rounding one of the
branchpoints and joining different edges of $L_+$ or $L_-$
respectively:
\begin{equation}\label{monodromy} {\bf
\Psi}\quad{\stackrel
{T_\pm}\longrightarrow}\quad\widetilde{{\bf\Psi}}= {\bf \Psi}
\cdot{\bf T}_\pm(w),\qquad
{\bf T}_\pm(w)={\bf I}-2\displaystyle{{\bf l}_\pm(w)\otimes{\bf
k}_\pm(w)\over ({\bf l}_\pm(w)\cdot{\bf
k}_\pm(w))}.
\end{equation}
It is remarkable, that these matrices, satisfying the identities
${\bf T}_\pm^2(w)\equiv{\bf I}$, are independent of the space-time
coordinates $\xi$, $\eta$. The structure (\ref{monodromy}) allows
to express ${\bf T}_\pm$ in terms of the four complex projective
vectors ${\bf k}_\pm(w)$ and ${\bf l}_\pm(w)$, but it was found
\cite{GA:1985} that the conditions (\ref{integral}) relate
unambiguously ${\bf l}_\pm(w)$ and ${\bf k}^\dagger_\pm(w)$ with the
same suffices. Therefore, all components of ${\bf T}_\pm$ are
determined completely by four scalar functions, which parametrise the
components of two projective vectors ${\bf k}_\pm(w)$:
\begin{equation}\label{mdata}
{\bf k}_\pm(w)=\{1\,,{\bf u}_\pm(w)\,,\,{\bf v}_\pm(w)\,\}
\end{equation}
The functions ${\bf u}_\pm (w)$, ${\bf v}_\pm (w)$, which domains
of holomorphicity are shown on \hbox{Fig.\ \ref{Fig_1}}, father are
referred to as monodromy data (we note here,
that for vacuum ${\bf v}_\pm(w)\equiv 0$), and  the described above
method for calculation of these functions for any solution of the
Ernst equations solves a direct problem of our monodromy transform.

\section{Inverse problem of the monodromy transform}

The statements, proved in \cite{GA:1985, GA:1988}, imply the
existence of two equivalent forms of the linear singular integral
equation, which are equivalent to the Ernst equations:
$$-\displaystyle{1\over \pi i}\vpint\limits_L\,{
[\lambda]_\zeta {\cal H}(\tau,\zeta)\over
\zeta-\tau}\,{\bf \varphi}(\xi,\eta,\zeta)\,d\,\zeta={\bf
k}(\tau),\quad
-\displaystyle{1\over \pi i}\vpint\limits_L\,{
[\lambda^{-1}]_\zeta {\cal H}(\zeta,\tau)\over \zeta-\tau}\,{\bf
\psi}(\xi,\eta,\zeta)\,d\,\zeta={\bf l}(\tau),$$
where $[\lambda]_\zeta$ is the jump of
$\lambda=\sqrt{(w-\xi)(w-\eta)/(w-\xi_0)(w-\eta_0)}$,
with $\lambda(\xi,\eta,\infty)=1$, at the
point $\zeta\in L$  and  ${\cal H}(\tau,\zeta)\equiv ({\bf k}(\tau)\cdot{\bf
l}(\zeta))$. Vector solutions ${\bf\varphi}(\xi,\eta,\tau)$,  ${\bf
\psi}(\xi,\eta,\tau)$ of these equations together with the
corresponding monodromy data vectors ${\bf k}$, ${\bf l}$ determine
completely the solution of our spectral problem and hence, of the
Ernst equations.

These equations admit equivalent regularizations by standard methods
(here - the "left regularization"), which lead to equivalent
equations of a (quasi-)Fredholm type:
\begin{equation}\label{fredholm} {\bf
\phi}(\tau)+\displaystyle{\int\limits_L}\,{\cal
F}(\tau,\zeta)\,{\bf\phi}(\zeta)\,d\,\zeta={\bf h}(\tau),\qquad {\bf
\psi}(\tau)+\displaystyle{\int\limits_L}\,\widetilde{\cal
F}(\tau,\zeta)\,{\bf\psi}(\zeta)d\,\zeta= \widetilde{\bf h}(\tau),
\end{equation}
with ${\bf \phi}(\tau)=-{\cal H}(\tau,\tau){\bf\varphi}(\tau)$
and the coefficients, determined by monodromy data:
$$\begin{array}{lll}
{\cal F}(\tau,\zeta)=[\lambda]_\zeta \displaystyle{1\over i\pi}
\vpint\limits_L\,{[\lambda^{-1}]_\chi\over
\chi-\tau}\,{\cal S}(\chi,\zeta)\,d\,\chi&&
{\bf h}(\tau)=\displaystyle{1\over i\pi}\vpint
\limits_L\,{[\lambda^{-1}]_\chi\over
\chi-\tau}\,\bf{k}(\chi)\,d\,\chi
\\ \widetilde{\cal
F}(\tau,\zeta)=-[\lambda^{-1}]_\zeta \displaystyle{1\over i\pi}
\vpint\limits_L\,{[\lambda]_\chi\over
\chi-\tau}\,{\cal S}(\zeta,\chi)\,d\,\chi&&
\widetilde{\bf h}(\tau)=-\displaystyle{1\over i\pi}
\vpint\limits_L\,{[\lambda]_\chi\over
\chi-\tau}\,{{\bf l}(\zeta)\over {\cal
H}(\zeta,\zeta)}\,d\,\chi,
\end{array}$$
where ${\cal S}(\tau,\zeta)=\displaystyle{{\cal H}(\tau,\zeta)-{\cal
H}(\zeta,\zeta)\over i\pi (\zeta-\tau){\cal H}(\zeta,\zeta)}$.
The local solution of each of the equations (\ref{fredholm}) for any
choice of monodromy data can be constructed by the known
iterative method:
\begin{equation}\label{series}
 \begin{array}{l}
{\bf \phi}(\tau)={\bf\phi}_0(\tau)+\sum\limits_{n=1}^\infty
\left({\bf\phi}_n(\tau)-{\bf\phi}_{n-1}(\tau)\right),\\[2ex]
{\bf\phi}_0(\tau)={\bf h}(\tau),\qquad
{\bf\phi}_n(\tau)={\bf h}(\tau)-\displaystyle{\int\limits_L}{\cal
 F}(\tau,\zeta){\bf\phi}_{n-1}(\zeta)\,d\,\zeta \end{array}
\end{equation}
A homogeneous convergency of these series, easily proven for local
solutions, when $\xi$, $\eta$ are close enough to their initial
values $\xi_0$, $\eta_0$, provides the existence and
uniqueness of solution of each of the discussed above
equivalent integral equations for arbitrary chosen
monodromy data. This means eventually that each of these equations
solves the inverse problem of the monodromy transform for
the Ernst equations.

\vspace*{0.25cm} \baselineskip=10pt{\small \noindent
This work was partly supported by the
British Engineering and
Physical Sciences Research Council and the Russian Foundation for
Basic Research Grants 99-01-01150, 99-02-18415.}

\end{document}